\documentclass{aa}
\usepackage{graphicx}

\def\jh{\mbox{$\rm (J-H)$}}
\def\hk{\mbox{$\rm (H-K_s)$}}
\def\mMJ{\mbox{$\rm (m-M)_J$}}
\def\mMo{\mbox{$\rm (m-M)_O$}}
\def\ebv{\mbox{$\rm E(B-V)$}}
\def\ejh{\mbox{$\rm E(J-H)$}}
\def\rc{\mbox{$\rm R_{core}$}}
\def\rl{\mbox{$\rm R_{lim}$}}
\def\ms{\mbox{$\rm M_\odot$}}
\def\ds{\mbox{$\rm d_\odot$}}
\def\dgc{\mbox{$\rm d_{GC}$}}
\def\jj{\mbox{$\rm J$}}
\def\hh{\mbox{$\rm H$}}
\def\ks{\mbox{$\rm K_s$}}

\def\mobs{\mbox{$\rm m_{obs}$}}

\def\kms{\mbox{$\rm km\,s^{-1}$}}
\def\tr{\mbox{$\rm t_{relax}$}}
\def\tcr{\mbox{$\rm t_{cross}$}}
\def\mas{\mbox{$\rm mas\,yr^{-1}$}}
\def\ee{\mbox{$\large\varepsilon$}}

\begin{document}

\title{Detection of \ks-excess stars in the 14\,Myr open cluster NGC\,4755}

\author{C. Bonatto\inst{1} \and E. Bica\inst{1} \and S. Ortolani\inst{2} \and 
B. Barbuy\inst{3}}

\offprints{Ch. Bonatto}

\institute{Universidade Federal do Rio Grande do Sul, Instituto de F\'\i sica, 
CP\,15051, Porto Alegre 91501-970, RS, Brazil\\
\email{charles@if.ufrgs.br, bica@if.ufrgs.br}
\mail{charles@if.ufrgs.br} 
\and
Universit\`a di Padova, Dipartimento di Astronomia, Vicolo dell'Osservatorio 2, 
I-35122 Padova, Italy\\
\email{ortolani@pd.astro.it}
\and
Universidade de S\~ao Paulo, Rua do Mat\~ao 1226, 05508-900, S\~ao Paulo, Brazil\\
\email{barbuy@astro.iag.usp.br}}

\date{Received --; accepted --}

\abstract{}{We derive the structure, distribution of MS and PMS stars and dynamical state of 
the young open cluster NGC\,4755. We explore the possibility that, at the cluster age, some MS 
and PMS stars still present infrared excesses related to dust envelopes and proto-planetary discs.}
{\jj, \hh\ and \ks\ 2MASS photometry is used to build CMD and colour-colour diagrams, 
radial density profiles, luminosity and mass functions. Field-star decontamination is applied to 
uncover the cluster's intrinsic CMD morphology and detect candidate PMS stars. Proper motions 
from UCAC2 are used to determine cluster membership.}
{The radial density profile follows King's law with a core radius $\rm\rc=0.7\pm0.1\,pc$
and a limiting radius $\rm\rl=6.9\pm0.1\,pc$. The cluster age derived from Padova isochrones is 
$\rm14\pm2\,Myr$. Field-star decontamination reveals a low-MS limit at $\rm\approx1.4\,\ms$. The core 
MF ($\chi=0.94\pm0.16$) is flatter than the halo's ($\chi=1.58\pm0.11$). NGC\,4755 contains 
$\rm\sim285$ candidate PMS stars of age $\rm\sim1 - 15\,Myr$, and a few evolved 
stars. The mass locked up in PMS, MS and evolved stars amounts to $\rm\sim1150\,\ms$. Proper motions 
show that \ks-excess MS and PMS stars are cluster members. \ks-excess fractions in PMS and MS stars 
are $\rm5.4\pm2.1\%$ and $\rm3.9\pm1.5\%$ respectively, consistent with the cluster age. 
The core is deficient in PMS stars, as compared with MS ones. NGC\,4755 hosts binaries in the halo 
but they are scarce in the core.} 
{Compared to open clusters in different dynamical states studied with similar methods, NGC\,4755 
fits relations involving structural and dynamical parameters in the expected locus for its age and 
mass. On the other hand, the flatter core MF probably originates from primordial processes related 
to parent molecular cloud fragmentation and mass segregation over $\rm\sim14\,Myr$. Star 
formation in NGC\,4755 began $\rm\approx14\,Myr$ ago and proceeded for about the same length of time. 
Detection of \ks-excess emission in member MS stars suggests that some circumstellar dust discs 
survived for $\rm\sim10^7\,yr$, occurring both in some MS and PMS stars for the age and spread 
observed in NGC\,4755.}


\titlerunning{\ks-excess stars in NGC\,4755}

\authorrunning{C. Bonatto et al.}

\maketitle

\section{Introduction}
\label{intro}

Before settling onto the zero-age main sequence (ZAMS), stars spend their early lives surrounded 
by optically thick material consisting of an infalling envelope and accretion disc that 
gradually disperse along the pre-main sequence (PMS) phase. The gas and dust disc and envelope 
(DDE) are reminiscent of the primordial gravitational collapse that formed the stars. Final stages 
of disc accretion may last as long as $\rm\sim10^7\,yr$ (White \& Hillenbrand \cite{WH05}) and the 
presence of such inner discs may be traced by observations in near-infrared bands. 

On theoretical grounds it is estimated that because of disc-depleting processes such as irradiation 
by the central star, viscous accretion and mass loss due to outflow, the median lifetime of optically
thick inner accretion discs may be as short as $\rm2 - 3\,Myr$ (Hillenbrand \cite{H05}). In fact, 
observations indicate that significant fractions of stars younger than 1\,Myr have already lost their 
discs. However, they also indicate that 8--16\,Myr old stars  may still 
retain their inner discs (e.g. Chen et al. \cite{Chen05}; Low et al. \cite{Low05}). Similar conclusions 
were reached by Armitage, Clarke \& Palla (\cite{Ar03}) who found that $\rm\sim30\%$ of the stars in 
young clusters lose their discs in less than 1\,Myr, while the remainder keep them for about 1-10\,Myr. 
Observational estimates of disc survival time-scales are important for planet formation theories 
(Brandner et al. \cite{Brand00}).

Circumstellar discs in T Tauri stars have been inferred from infrared excesses originated in dust-heated 
emission (Hillenbrand \cite{H05}; Ortolani et al. \cite{OBBM05} and references therein). 
The current consensus is that near-infrared excesses correlate inversely with stellar age. Near-infrared 
colour-colour diagrams (2-CD) involving JHKL or \ks-bands have been widely used to identify 
infrared-excess stars and candidate circumstellar discs in young clusters (Hillenbrand \cite{H05}; 
Lada \& Adams \cite{LaAd92}). Thus, detection of DDEs around stars in young clusters with accurate age 
determination is crucial to observationally constrain dust disc time-scales. 

In addition to the disc-lifetime problem, stars in young open clusters
do not all have the same age, but formed along some period of time (e.g. Sagar \& Cannon
\cite{Sagar95} and references therein). Inferences on star-formation time-scales may be 
made by comparing the age derived from the bulk of main sequence (MS) stars with that 
implied from the observed distribution of PMS stars. This in turn may provide  
constraints on the time-scale of the parent molecular cloud fragmentation.

Another issue associated with molecular cloud fragmentation is the apparent universality of
the initial mass function (IMF). Observations of star-forming regions in molecular clouds, 
rich star clusters and Galactic field suggest that the IMF is similar in these very 
different environments (Kroupa \cite{Kroupa2002}). This suggests that the initial 
distribution of stellar masses depends on the process of molecular cloud fragmentation.
Fragmentation, in turn, should produce similar IMFs despite very different initial conditions, a 
not yet understood physical process (Kroupa \cite{Kroupa2002}). 

Detailed analysis of stellar density structure and stellar-mass distribution in young star 
clusters may shed light on these issues. Rich young clusters at the post embedded phase 
represent ideal targets because they contain large numbers of stars with different masses and provide 
uniformity in parameters such as age, chemical composition, distance and reddening. Assigning
accurate ages to such clusters by means of theoretical isochrones is feasible because
of evolved-star and/or PMS features in observed colour-magnitude diagrams (CMDs).

Recently, the structure and stellar distribution of the very young embedded open cluster NGC\,6611 
(age $\rm\approx1.3\pm0.3\,Myr$) was analyzed with near-infrared CMDs and 2-CDs (Bonatto, 
Santos Jr. \& Bica \cite{BSJB05}). Despite the young age, NGC\,6611 presents signs of dynamical 
evolution, such as a mass function (MF) in the cluster core significantly flatter than the halo's. 
They suggested that such MF flattening in the core is probably associated with the parent molecular 
cloud fragmentation, where more massive proto-stars are preferentially located in the central parts, 
which agrees with Kroupa (\cite{Kroupa2004}) in the observed degree of mass segregation 
in clusters younger than a few Myr. In addition, NGC\,6611 was shown to contain a significant number 
of PMS stars with ages in the range 0.2--4\,Myr. Some of these PMS stars present infrared excess 
emission.

The rich open cluster NGC\,4755, with an age in the range 10 -- 20\,Myr (e.g. Sanner et al. 
\cite{SBWG01}; Nilakshi et al. \cite{Nilakshi2002}) and evidence of hosting PMS stars (e.g. Sagar \& 
Cannon \cite{Sagar95}), is another candidate to search for PMS stars and infrared excess 
emission. We will follow similar procedures as for the comparison open cluster NGC\,6611.

Our goals with the present work are {\em (i)} to carry out a detailed analysis of the density structure 
of NGC\,4755, {\em (ii)} analyze the core, halo and overall cluster MFs, {\em (iii)} derive properties 
of MS and PMS stars and cluster parameters, {\em (iv)} infer information on the dynamical state, and 
{\em (v)} search for infrared-excess stars at such ages. For 
spatial and photometric uniformity, we employ \jj, \hh\ and \ks\ 
2MASS\footnote {The Two Micron All Sky Survey, All Sky data release (Skrutskie et al. \cite{2mass1997}), 
available at {\em http://www.ipac.caltech.edu/2mass/releases/allsky/}} photometry. The 2MASS Point 
Source Catalogue (PSC) is uniform reaching relatively faint magnitudes covering nearly all the sky, 
allowing a proper background definition for clusters with large angular sizes (e.g.
Bonatto, Bica \& Santos Jr. \cite{BBS2005}; Bonatto, Bica \& Pavani \cite{BBP2004}). 

This paper is organized as follows. In Sect.~\ref{n4755} we review previous results on NGC\,4755. 
In Sect.~\ref{2mass} we present the 2MASS data, subtract field-star contamination, derive fundamental
cluster parameters, discuss candidate PMS stars and analyze the cluster density structure. In Sect.~\ref{MF}
we derive luminosity functions (LFs) and MFs and discuss stellar content properties. In Sect.~\ref{dyna} we 
compare NGC\,4755 with open clusters in different dynamical states. Concluding remarks are given
in Sect.~\ref{Conclu}. 

\section{The post-embedded open cluster NGC\,4755}
\label{n4755}

\begin{figure}
\resizebox{\hsize}{!}{\includegraphics{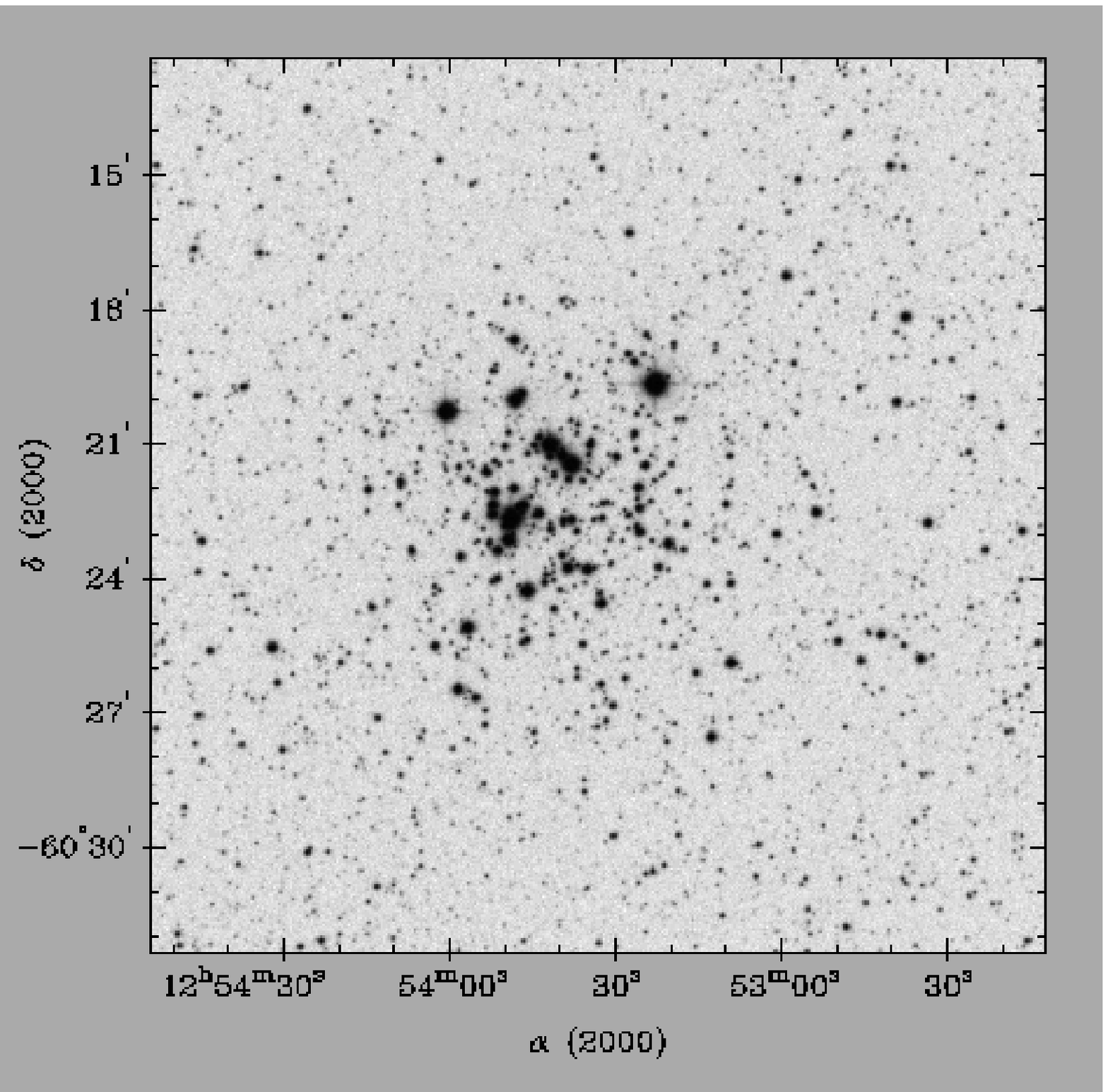}}
\caption[]{DSS B image of NGC\,4755 covering $\rm20\arcmin\times20\arcmin$. }
\label{fig1}
\end{figure}

NGC\,4755, also known as $\kappa$ Crucis or {\em Herschell's Jewel Box}, is a prominent young 
open cluster in the southern hemisphere. The bright stars characterizing this cluster can be
seen in the DSS\footnote{Extracted from the Canadian Astronomy Data Centre (CADC), at \em 
http://cadcwww.dao.nrc.ca/} B image (Fig.~\ref{fig1}) covering $\rm20\arcmin\times20\arcmin$. 
This cluster contains the red supergiant (SG) $\rm CPD-59^\circ4547$ of spectral type M2Iab and 
$\rm V=7.45$ and $\rm (B-V)=2.22$ (Dachs \& Kaiser \cite{DK84}). They also classified 5 blue 
SGs with $\rm 5.7<V< 9.1$ in NGC\,4755. They detected several extremely
red stars in the cluster field by means of UBV photometry. They have near-infrared 
counterparts in the present study, seen especially in a 2-CD (Sect.~\ref{PMS}).

No evidence of gas emission or dust filaments appears in this B DSS image (such features are 
almost totally absent in the R XDSS image) of NGC\,4755, which we refer to as a post-embedded young 
open cluster.

Lyng\aa\ (\cite{Lynga87}) classified NGC\,4755 as Trumpler type I3r, with angular diameter
$\rm D=10\arcmin$, and estimated a MS turn-off age in the range $\rm7-24\,Myr$.

Sagar \& Cannon (\cite{Sagar95}) using deep UBVRI CCD photometry estimated a distance from 
the Sun $\rm\ds\approx2.1\pm0.2\,kpc$. They found variable 
reddening following a normal extinction law with an average value $\rm\ebv=0.41$ and 
a differential reddening range $\rm\Delta\ebv=0.05$, a picture typical of a post-embedded
young open cluster, where molecular and dust clouds are dissipating. This is 
consistent with the near absence of dust filaments in NGC\,4755 (Fig.~\ref{fig1}). They 
suggested that star-formation in the parent molecular cloud might have proceeded for at least 
$\rm6-7\,Myr$, and massive stars ($\rm m>10\,\ms$) formed $\rm\sim4\,Myr$ before the bulk of 
the low-mass stars ($\rm m<2\,\ms$). They also detected PMS stars with ages in the range 
$\rm3-10\,Myr$.

Based on BVI CCD photometry and astrometric data, Sanner et al. (\cite{SBWG01}) derived 
$\rm\ds\approx2.1\pm0.2\,kpc$, $\ebv=0.36\pm0.02$, an age of $\rm10\pm5\,Myr$, and an IMF 
with slope $\chi=1.68\pm0.14$ for stars in the mass range $\rm1.2\leq m(\ms)\leq13.6$. They
derived a solar metallicity.
 
Using UBV CCD observations Tadross et al. (\cite{Tad2002})
derived for NGC\,4755 a colour excess $\ebv=0.38$, $\ds\approx1.8$\,kpc, age $\approx34$\,Myr, 
linear diameter $\rm D=5.4$\,pc, Galactocentric distance $\dgc=7.66$\,kpc, number of member 
stars $\rm N_*=364$, and cluster mass of $\rm m=682\,\ms$.

Nilakshi et al. (\cite{Nilakshi2002})  using photometry from Palomar Observatory Sky Survey 
I plates found $\ds\approx2.1$\,kpc, $\dgc=7.6$\,kpc, age 
$\approx20$\,Myr, core radius $\rm\rc=0.63\pm0.10\,pc$ and core stellar density $\rm\rho_c=21.1
\pm3.8\,stars\,pc^{-2}$, halo radius $\rm R_h=6.1\,pc$ and halo stellar density $\rm\rho_h=7.4
\pm0.3\,stars\,pc^{-2}$.

Piskunov et al. (\cite{Piskunov04}) using BV CCD photometry derived an age of $\rm16\pm1\,Myr$, 
IMF slope $\chi=1.4\pm0.3$ and a star-formation spread of $\rm1\pm1\,Myr$.

In the WEBDA\footnote{\em http://obswww.unige.ch/webda} open cluster database 
(Mermilliod \cite{Merm1996}) the central coordinates of NGC\,4755 are (J2000) $\alpha=12^h53^m39^s$, 
and $\delta=-60^\circ21\arcmin42\arcsec$. However, the radial density profile (Sect.~\ref{struc}) for 
these coordinates presented a dip at $\rm R=0\arcmin$. We searched for a new center by examining 
histograms for the number of stars in 0.5\arcmin\ bins of right ascension and declination. The
coordinates that maximize the density of stars at the center are (J2000) $\alpha=12^h53^m33.1^s$, 
and $\delta=-60^\circ22\arcmin21.0\arcsec$, corresponding to $\ell=303.193^\circ$ and $b=+2.498^\circ$. 
In what follows we refer to these optimized coordinates as the center of NGC\,4755. WEBDA gives 
$\ebv=0.386$, $\ds=1.98$\,kpc and $\rm age=16\,Myr$.

\section{Cluster parameters from 2MASS data}
\label{2mass}

VizieR\footnote{\em http://vizier.u-strasbg.fr/viz-bin/VizieR?-source=II/246} was used to 
extract \jj, \hh\ and \ks\ 2MASS photometry in a circular area with radius 
$\rm R=50\arcmin$ centered on the optimized coordinates of NGC\,4755 (Sect.~\ref{n4755}). As a
photometric quality constraint, the extraction was restricted to stars brighter than the 99.9\% Point 
Source Catalogue Completeness Limit\footnote{Following the Level\,1 Requirement, according to 
{\em\tiny http://www.ipac.caltech.edu/2mass/releases/allsky/doc/sec6\_5a1.html }}, $\jj=15.8$, 
$\hh=15.1$\ and $\ks=14.3$, respectively. For reddening transformations we use the 
relations $\rm A_J/A_V=0.276$, $\rm A_H/A_V=0.176$ and $\rm A_{K_S}/A_V=0.118$ (Dutra, Santiago
\& Bica \cite{DSB2002}), assuming a total-to-selective absorption ratio $\rm R_V=3.1$. 

\begin{figure} 
\resizebox{\hsize}{!}{\includegraphics{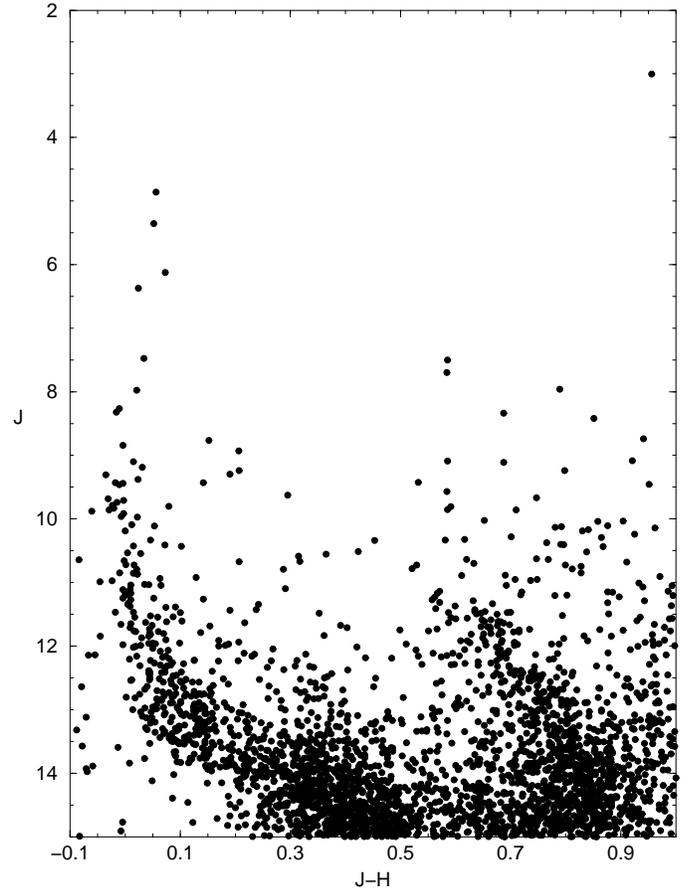}}
\caption[]{$\jj\times\jh$ CMD of the central 10\arcmin\ of NGC\,4755 (blue sequence). Field-star 
contamination, mostly by disc stars, is conspicuous. The extension of the MS to $\rm\jj<14$
appears to be a field-star contamination artifact.}
\label{fig2}
\end{figure}

The CMD of the central 10\arcmin\ of NGC\,4755 is given in Fig.~\ref{fig2}. Reflecting the young age, 
a prominent, nearly vertical MS is present, together with a single bright star at
$\rm\jj\approx3$ and $\rm\jh\approx0.95$ (the red SG $\rm CPD-59^\circ4547$). Because of 
its low latitude, field stars (mostly disc) contaminate the CMD, particularly at faint magnitudes 
and red colours mimicking a MS extending to sub-solar mass stars, which is not compatible
with a 10 --20\,Myr old cluster. 

\subsection{Field-star decontamination}
\label{FSD}

\begin{figure} 
\resizebox{\hsize}{!}{\includegraphics{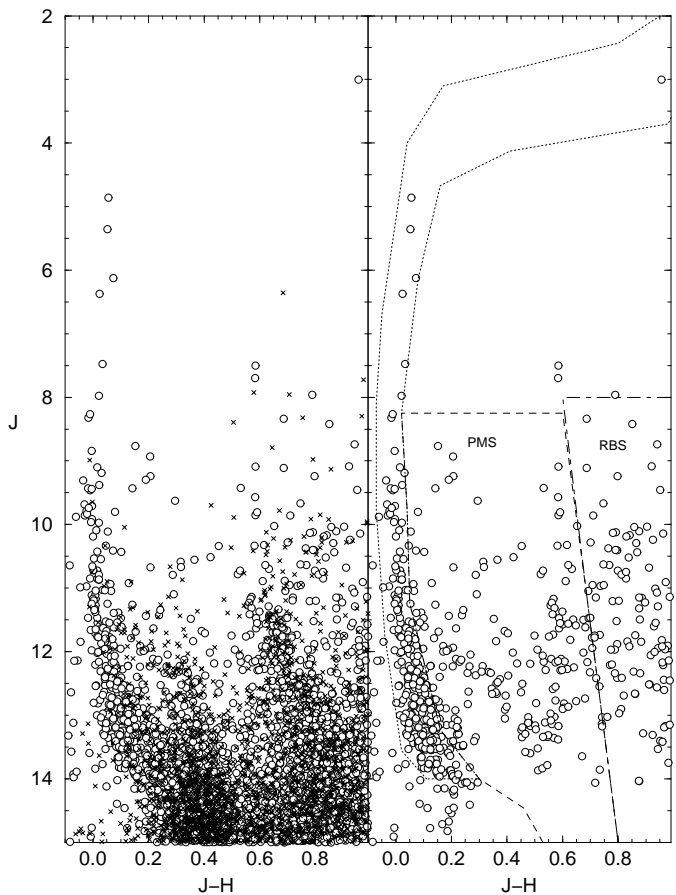}}
\caption[]{Left panel: field stars ('x') are superimposed on the observed CMD of the central
10\arcmin\ of NGC\,4755. Extraction areas match. Right panel: field-star decontaminated CMD. 
Colour filters used to isolate MS (dotted lines), candidate-PMS (dashed) and very red (dot-dashed) 
stars are shown.}
\label{fig3}
\end{figure}

The relative density and colour-magnitude distribution of the field-star contamination can be 
evaluated in the left panel of Fig.~\ref{fig3} where we superimpose on the observed CMD of the 
central 10\arcmin\ the corresponding (same area) field-star contribution. Most of the 
faint ($\rm\jj\leq14$) and red ($\jh\geq0.5$) stars probably are contaminant field stars.
To retrieve the intrinsic cluster-CMD morphology we use a field-star decontamination procedure
previously applied in the analysis of low-contrast (Bica \& Bonatto \cite{LowC05}) and young 
embedded (Bonatto, Santos Jr. \& Bica \cite{BSJB05}) open clusters. As the offset field we take the 
region located at $\rm30\arcmin\leq R\leq50\arcmin$. This area is large enough to produce 
field-star statistical representativity ($\rm\approx78\,000\,stars$), both in magnitude and 
colours.

Based on the spatial number-density of stars in the offset field, the decontamination procedure 
estimates the number of field stars that within the $\rm1\sigma$ level are expected to be present 
in the cluster field. The observed CMD is divided in colour/magnitude cells from which stars 
are randomly subtracted in a number consistent with the expected number of field stars in the same 
cell. Dimensions of the colour/magnitude cells can be subsequently changed so that the total 
number of stars subtracted throughout the whole cluster area matches the expected one, at the 
$\rm1\sigma$ level. Since field stars are taken from an external region of fixed dimension, 
corrections are made for differences in solid angle between cluster and offset field regions. 
This procedure can be applied to the whole cluster and internal regions as well. Because it 
actually excludes stars from the original files - thus artificially changing both the radial 
distribution of stars and LFs - we use field-star decontamination only to uncover the intrinsic 
CMD morphology and build 2-CDs. 

 The number of observed stars in the central 10\arcmin\ field of NGC\,4755 is 5510. Since the 
field-star ($\rm30\arcmin\leq R\leq50\arcmin$) density is $\rm13.6\pm0.15\,stars\,(\arcmin)^{-2}$ the 
expected number of cluster stars in that region amounts to $1246\pm99$ ($22.6\%$), while field stars
are $4264\pm65$ ($\rm77.4\%$). These numbers were obtained with colour/magnitude cells of
dimensions 0.01 and 0.025, respectively.

The field-star decontaminated CMD of the central 10\arcmin\ is in the right panel of 
Fig.~\ref{fig3}. As expected, most of the faint and red stars were eliminated by the 
decontamination procedure. The decontaminated CMD morphology presents an extended, nearly-vertical 
MS with the faint-limit at $\rm\jj\approx14$. It suggests the presence of a significant population 
of candidate PMS stars at $\rm0.2\leq\jh\leq0.8$, and a number of very red stars at $\rm\jh\geq0.8$ 
that are probably red background stars (RBS - Sect.~\ref{PMS}). Indeed, the presence of background 
disc stars is expected in the direction of NGC\,4755, since we are observing at $\rm\ell\approx303^\circ$ 
and low latitude. In Fig.~\ref{fig3} we show the colour-magnitude filters used to isolate 
MS/evolved, candidate PMS and RBS stars. The filters were defined based on the distribution of the 
decontaminated star sequences as compared with theoretical MS and PMS tracks (Fig.~\ref{fig4}). 
 The MS filter follows the 14\,Myr Padova isochrone (Sect.~\ref{age}) allowing for
photometric uncertainties, while the PMS filter describes the colour/magnitude space 
covered by the 0.1 -- 20\,Myr PMS tracks (Sect.~\ref{PMS}). The MS/evolved, candidate-PMS and RBS 
colour-filters applied to the observed photometry of the central 10\arcmin\ field (left panel of
Fig.~\ref{fig3}) select 346, 1131 and 953 stars, respectively. Applied to the field-star decontaminated 
photometry (right panel of Fig.~\ref{fig3}), the filters select 277 ($80\pm6\%$ of the observed ones), 
285 ($25\pm2\%$) and 204 ($21\pm2\%$) stars, respectively for MS/evolved, candidate-PMS and RBS stars.

\subsection{Cluster age and distance from the Sun}
\label{age}

Cluster age is derived with solar-metallicity Padova isochrones (Girardi et al. 
\cite{Girardi2002}) computed with the 2MASS \jj, \hh\ and \ks\ filters\footnote{\em\tiny 
http://pleiadi.pd.astro.it/isoc\_photsys.01/isoc\_photsys.01.html. 2MASS transmission 
filters produced isochrones very similar to the Johnson ones, with differences of at most 
0.01 in \jh\ (Bonatto, Bica \& Girardi \cite{BBG2004})}.

\begin{figure} 
\resizebox{\hsize}{!}{\includegraphics{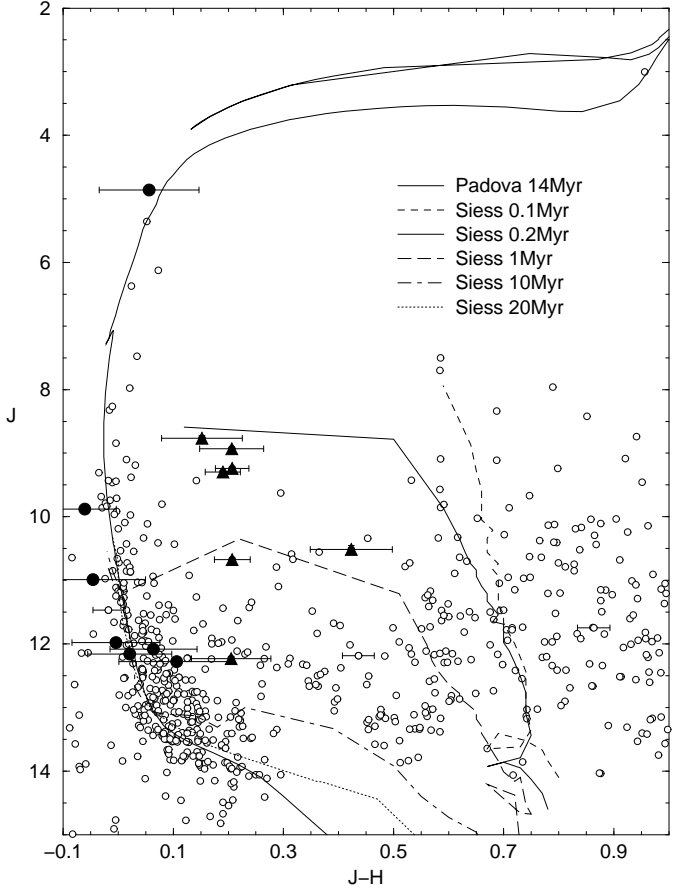}}
\caption[]{Isochrone fit of the decontaminated CMD of the central 10\arcmin\ of NGC\,4755. The 
best fit was obtained with the 14\,Myr solar-metallicity Padova isochrone, $\rm\ejh=0.07$ 
and $\rm(m-M)_J=11.5$. PMS tracks (0.1, 0.2, 1, 10 and 20\,Myr) are plotted following the same 
$\rm(m-M)_J$ and \ejh\ solution. \ks-excess MS (filled circles) and PMS (filled triangles) stars 
are shown along with the individual error bars (Sect.~\ref{KSEF}). To avoid cluttering, only 
representative error bars are shown for non-\ks-excess stars.} 
\label{fig4}
\end{figure}

It is usually difficult to derive age by means of CMDs for young star clusters because of 
nearly-vertical MS and scarcity of evolved stars. The lack of observational CMD constraints 
allows several age solutions. In the field-star decontaminated CMD of NGC\,4755 the red SG 
together with the rather well-defined low-MS reaching $\rm\jj\approx14$ constrain the age to 
$\rm14\pm2\,Myr$. Parameters derived from the isochrone fit are the observed distance modulus 
$\rm\mMJ=11.5\pm0.1$ and colour excess $\rm\ejh=0.07\pm0.01$, converting to $\rm\ebv=0.22\pm0.03$. 
This age solution is plotted in Fig.~\ref{fig4}. With these parameters the absolute distance modulus 
is $\rm\mMo=11.3\pm0.1$, resulting in $\rm\ds=1.8\pm0.1\,kpc$. The Galactocentric distance of 
NGC\,4755 is $\dgc=7.2\pm0.2$\,kpc, using 8.0\,kpc as the Sun's distance to the Galactic center 
(Reid \cite{Reid93}). MS stars are restricted to the mass range $\rm1.4\leq m(\ms)\leq13.5$.

\subsection{Cluster structure}
\label{struc}

Cluster structure was inferred by means of the radial density profile (RDP), defined as the projected
number-density of cluster stars around the center.  The MS/evolved RDP was built with stars selected
after applying the respective colour-magnitude filter shown in Fig.~\ref{fig3}. Within uncertainties, 
the filter describes the cluster's intrinsic CMD morphology from the upper MS/evolved stars to the 
turn-on region at $\rm J\approx14$. This procedure allowed us to exclude most of the background field 
objects. The use of colour-magnitude 
filters to discard stars with discordant colours was previously applied in the analysis of the open 
clusters M\,67 (Bonatto \& Bica \cite{BB2003}), NGC\,188 (Bonatto, Bica \& Santos Jr. \cite{BBS2005}) 
and NGC\,3680 (Bonatto, Bica \& Pavani \cite{BBP2004}). To avoid oversampling near the center and 
undersampling for large radii, the RDP was built counting stars in rings with radius 
$\rm\Delta R=0.5\arcmin$ for $\rm 0\leq R(\arcmin)<5$, $\rm\Delta R=1\arcmin$ for $\rm 5\leq 
R(\arcmin)<10$, $\rm\Delta R=2\arcmin$ for $\rm 10\leq R(\arcmin)<20$ and $\rm\Delta R=4\arcmin$ for 
$\rm R\geq20\arcmin$. The residual background level corresponds to the average number of stars in the 
region located at $\rm 28\arcmin\leq R\leq 46\arcmin$, resulting in $\rm\sigma_{bg}=0.203\pm0.007\,stars\,
(\arcmin)^{-2}$.

\begin{figure} 
\resizebox{\hsize}{!}{\includegraphics{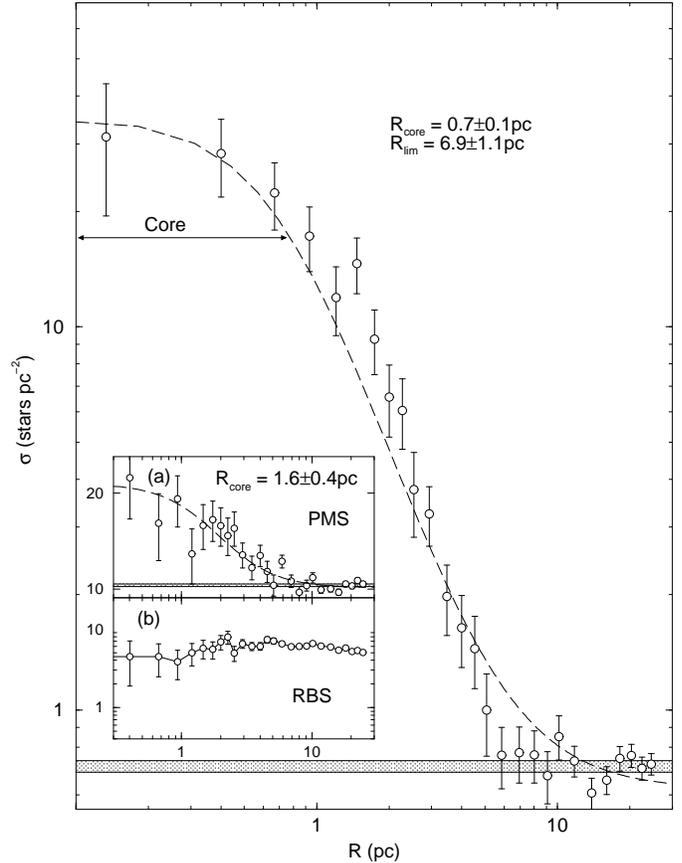}}
\caption[]{Radial density profile (MS) of NGC\,4755. Dashed line: two-parameter King profile;
core size is indicated. Shaded region: stellar background level. Insets: RDP of PMS 
(panel a) and RBS (panel b) stars. }
\label{fig5}
\end{figure}

Fig.~\ref{fig5} shows the MS/evolved stars' RDP. For absolute comparison between clusters the radius 
scale was converted to parsecs and the number-density of stars to $\rm stars\,pc^{-2}$\ using the 
distance derived in Sect.~\ref{age}. The high statistical significance of the RDP is reflected in the 
$1\sigma$\ Poisson error bars. 

Structural parameters of NGC\,4755 were derived by fitting the two-parameter King (\cite{King1966a}) 
surface density profile to the background-subtracted RDP. King's model describes the 
intermediate and central regions of normal clusters (King \cite{King1966b}; Trager, King \& Djorgovski 
\cite{TKD95}). The fit was performed using a nonlinear least-squares fit routine that uses errors as 
weights. The best-fit solution (Fig.~\ref{fig5}) is superimposed on the observed RDP. Parameters 
derived are King's background-subtracted central density of stars 
$\rm\sigma_{0K}=9.9\pm1.4\,stars\,(\arcmin)^{-2}=34.8\pm5.1\,stars\,pc^{-2}$, and core radius 
$\rm\rc=1.38\pm0.15\arcmin=0.74\pm0.08\,pc$. Considering fluctuations of the observed RDP with respect 
to the background the cluster extends up to a limiting radius $\rm\rl=13\pm2\arcmin=6.9\pm1.1\,pc$. 
\rl\ describes where the observed RDP merges into the background and for practical purposes 
the bulk of the cluster stars are contained within $\rl$. The present values of \rc\ and \rl\ agree at 
the $\rm1\sigma$ level with those of Nilakshi et al. (\cite{Nilakshi2002}).

To determine membership we also build RDPs for stars located in the PMS and RBS CMD regions (Fig.~\ref{fig3}). 
Both sets of stars were isolated by the respective colour-magnitude filters.  Despite large 
errors, the RDP of PMS stars is well fitted by a King's profile with $\rm\sigma_{0K}=11.3\pm2.6\,stars\,pc^{-2}$ 
and core radius $\rm\rc=1.6\pm0.4\,pc$. This value of $\rm\rc$ differs $\rm\sim2\sigma$ from that derived 
with MS/evolved stars, while the central number-density of PMS stars is $\rm\sim1/3$ of that of the MS stars. 
Both the difference in core radius and number-density with respect to MS stars suggest a deficiency of PMS 
stars in the core (Sect.~\ref{DPMS}). The PMS limiting radius $\rm\rl=8\pm1\,pc$ agrees with the MS/evolved 
one. We conclude that cluster membership of the candidate-PMS stars is statistically indicated by their RDP 
(panel a) that presents a similar limiting radius as that of the MS/evolved stars and is represented well by 
King's profile.

On the other hand, the nearly flat RDP of RBS stars (panel b) represents that expected from the background. 

King's profile provides a good fit to the stellar RDP of NGC\,4755, particularly for MS/evolved stars. 
Since it follows from an isothermal (virialized) sphere, the close similarity of the stellar 
RDP with a King profile may suggest that the internal structure of NGC\,4755 (particularly the core) 
has reached some level of energy equipartition after $\rm\sim14\,Myr$. Part of this effect may be linked 
to molecular cloud fragmentation (Sect.~\ref{dyna}), similarly to NGC\,6611 
(Bonatto, Santos Jr. \& Bica \cite{BSJB05}). 

\subsection{Candidate pre-main sequence stars}
\label{PMS}

A significant number of stars to the right of the MS remains in the decontaminated CMD (Figs.~\ref{fig3} 
and \ref{fig4}). In young open clusters this CMD region is usually occupied by PMS stars. To characterize 
these stars in terms of age and mass we use PMS tracks of Siess, Dufour \& Forestini (\cite{Siess2000}) 
with ages 0.1, 0.2, 1, 10 and 20\,Myr (the 0.1\,Myr track is used to mark off the red PMS 
threshold). To these tracks we apply the $\rm(m-M)_J$ and \ebv\ values derived from the Padova 
isochrone fit (Sect.~\ref{age}). The PMS tracks are given in Fig.~\ref{fig4}. Candidate PMS stars are 
contained within the 1\,Myr and 20\,Myr tracks, and the turn-on points of the 10 and 20\,Myr PMS tracks 
coincide with the MS low-mass limit ($\rm\approx1.4\,\ms$). Within uncertainties, cluster age -
or at least the age of the earliest star formation event in NGC\,4755 - implied by the oldest PMS turn-on 
points is consistent with that derived with the Padova isochrone. The coincidence in age of the oldest 
PMS turn-on with the MS, together with the PMS age spread, indicates that star formation in NGC\,4755 began 
$\rm\sim14\,Myr$ ago and proceeded for about the same time.

\begin{figure} 
\resizebox{\hsize}{!}{\includegraphics{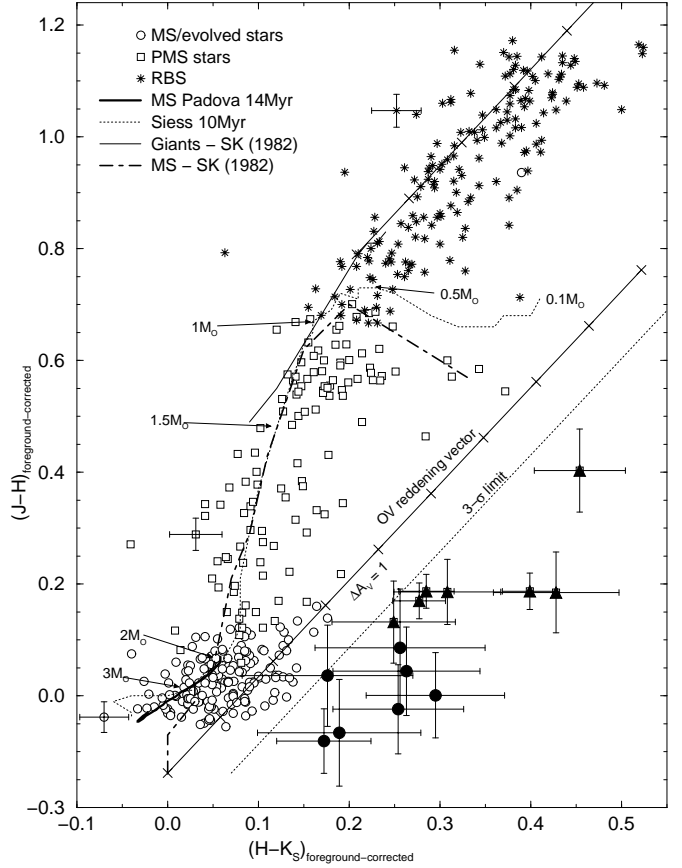}}
\caption[]{Foreground reddening-corrected ($\rm E(J-H)=0.07$ and $\rm E(H-\ks)=0.04)$) 2-CD 
of the field-star decontaminated 10\arcmin\ central region of NGC\,4755. Photometry was restricted 
to $\jj\leq13$. Circles: MS/evolved stars. Squares: PMS stars. Black stars: RBS. Heavy-solid line: 
14\,Myr Padova MS, restricted to $\rm2.0\leq m(\ms)\leq13.5$. Dotted line: 10\,Myr PMS track; representative 
PMS stellar masses are indicated. MS (dot-dashed line) and giant (solid) tracks are from Schmidt-Kaler 
(\cite{SK82}). Extinction vectors are shown with $\rm\Delta A_V=1\,mag$ subdivisions. Mean error bars 
in colour are shown for MS, PMS and RBS stars. \ks-excess MS (filled circles) and PMS (filled triangles) stars 
are to the right of the $\rm3\sigma$-limit line (light-dotted).}
\label{fig6}
\end{figure}

Properties of stars in the field of NGC\,4755 are also inferred by means of the 2-CD $\rm (H-\ks)\times(J-H)$, 
built with foreground reddening-corrected photometry of the central 10\arcmin\ region (Fig.~\ref{fig6}). To 
minimize photometric uncertainties and/or spurious detections we restricted the analysis to stars brighter 
than $\rm J=13$, taken from the field-star decontaminated photometry (Sect.~\ref{FSD}).  At this magnitude 
range we are probing stars more massive than $\rm\approx2\,\ms$. Mean photometric errors for MS, PMS and RBS 
stars are shown. For comparison we plot the 14\,Myr Padova isochrone restricted to the magnitude range 
$\rm\jj=13$ to the turnoff ($\rm\approx2.0 - 13.5\,\ms$), the 10\,Myr PMS track and the MS and giants loci 
of Schmidt-Kaler (\cite{SK82}). Schmidt-Kaler's MS includes stars more massive than $\rm\approx0.2\,\ms$. 
Except for low-mass stars the 10\,Myr PMS track and Schmidt-Kaler's MS are coincident. To characterize 
colour excesses we use the OV/late dwarfs reddening vector that is based on Schmidt-Kaler's loci.
Reddening vectors follow the relation $\rm E(J-H)=1.72\times E(H-\ks)$ (Sect.~\ref{2mass}). As expected,
reddening-corrected MS stars distribute about the Padova isochrone (and Schmidt-Kaler's MS for $\rm m>2\,\ms$), 
but with a redward bias probably resulting from uncorrected photospheric reddening and some amount of \ks-excess
emission. Most of the candidate PMS stars follow the 10\,Myr track at the $\rm\approx1\sigma$ level and are 
bluer than the OV reddening vector. A few MS and candidate-PMS stars present \ks-excess. 

Differential reddening cannot account for the large dispersion in colour observed in the
CMD and 2-CDs (Figs.~\ref{fig4} and \ref{fig6}, respectively), since throughout the field 
of NGC\,4755 it amounts to $\rm\Delta\ebv=0.05$ (Sagar \& Cannon \cite{Sagar95}).

As a further check on the reddened disc nature of the RBS stars we fitted a straight line to 
their colour-colour distribution. The resulting slope matches that of the reddening vector at 
the $\rm1\sigma$ level.

The presence of member-PMS stars of different ages indicates that star formation in NGC\,4755 
did not occur as a single event about 14\,Myr ago. Instead, massive stars and most of those more
massive than $\rm\sim2\,\ms$ probably formed in the early collapse, but the time-scale
for the formation of low-mass stars seems to be $\rm\sim10^7\,yr$. 

\subsubsection{\ks-excess emission stars}
\label{KSEF}

Disc strength as measured by excesses in intrinsic $\rm(H-K)$ or $\rm(K-L)$ (or \ks) colours 
depends on stellar mass and radius, and disc properties such as accretion rate and geometry 
(D'Alessio et al. \cite{DA99}). In particular, $\rm(H-\ks)$ is sensitive to discs with 
high accretion rates and low inclinations (Hillenbrand \cite{H05}). 

In the present case we quantify disc strength by the fraction of stars with intrinsic \ks-excess 
($\rm f_{K_s}$). Intrinsic colour excesses are obtained by subtracting the foreground reddening 
from the observed colours, and applying a correction for the underlying stellar photosphere. Since 
we do not have photospheric corrections for each MS and PMS star we take as intrinsic \ks-excess 
stars those with the foreground-reddening corrected colour \hk\ redder than the OV reddening vector 
by $\rm\Delta(H-\ks)\approx3\sigma\approx0.06\,mag$. According to this condition 
defined by the envelope of the reddest PMS stars (Fig.~\ref{fig6}), 14 \ks-excess stars are detected 
in this 14\,Myr old cluster. Half of them belong to the MS and the remainder to the PMS. We emphasize 
that these \ks-excess stars were taken from a field-star decontaminated sample of 2MASS stars brighter 
than $\rm J=13$. This restriction also precludes spurious measurements. Individual 
errors for these stars are shown in Fig.~\ref{fig6}. Although with errors larger than the mean, the 
loci occupied by the \ks-excess stars are consistent with the selection criterion adopted above.

The \ks-excess stars above do not occupy the classical disc locus of unreddened T Tauri stars; instead
they are somewhat bluer, particularly in \jh. However, theoretical models of classical full discs (e.g. 
Lada \& Adams \cite{LaAd92}) predict colour excesses that cover this region. These models extend to 
higher photospheric temperatures (to include Ae/Be stars), which in turn produce bluer colours. 
 Alternatively, part of the excess blue colour might be related to disc evolution, since as
discs age the inner regions are believed to dissipate more rapidly than the outer parts, creating 
central clearings. The role of central clearings in producing colours bluer than those of 
T Tauri stars should be investigated by modelling disc evolution.
To further investigate the blue \jh\ colours of the \ks-excess stars in NGC\,4755 we compared 2MASS 
and Kenyon \& Hartmann's (\cite{KH95}) IR photometry of stars in the Taurus-Auriga region. We found 
that the number of stars with blue colours in 2MASS is not larger than in Kenyon \& Hartmann's (\cite{KH95}) 
photometry, which indicates that the probability of blue colours in 2MASS is similar to other standard 
photometries. Also, Taurus-Auriga stars with colours below those of 
classical T Tauri stars present millimeter-emission from wide discs (Andrews \& Williams \cite{AW05}).

Despite the present photometric quality constraints we cannot completely rule out observational 
errors and/or anomalous colours. Variability, however, is not a problem, since the J, H and \ks\ 
photometry was obtained on the same night, as quoted in 2MASS.

One possible conclusion is that most of the \ks-excess stars in NGC\,4755 have a peculiar, hot full 
disc, and that at least the 4 bright stars (at $\jh\approx0.19$ and $\jj\approx9$ - Fig.~\ref{fig4}) 
could be Ae/Be stars, being also quite luminous. Most of the
considerations on the relation of circumstellar discs with colour have been based on observations 
of star clusters significantly younger than the 14\,Myr old NGC\,4755.

Bearing in mind the above caveats we estimate \ks-excess fractions of $\rm f_{K_s}=3.9\pm1.5\%$ and 
$\rm f_{K_s}=5.4\pm2.1\%$ for MS and PMS stars.  Uncertainties were estimated 
assuming statistical Poisson distributions for the field-star decontaminated MS and PMS populations.
Such low values of $\rm f_{K_s}$ are 
consistent with the age of NGC\,4755, according to the $\rm age\times f_{K_s}$ diagram (Hillenbrand 
\cite{H05}). A less conservative criterion for selecting \ks-excess stars, extending bluewards to 
the OV reddening vector, could double the $\rm f_{K_s}$ of MS stars, but would not change the value for 
PMS stars. The fractions would still be consistent with the cluster age. 

MS and PMS \ks-excess emission stars detected in the 2-CD (Fig.~\ref{fig6}) are identified
in the field-star decontaminated CMD (Fig.~\ref{fig4}). The \ks-excess MS stars are
distributed about the expected MS locus, including 1 massive star.

We conclude that the field-star decontaminated CMD and 2-CD present some MS and PMS stars with 
\ks-excess emission. If the bulk of present-day MS stars in NGC\,4755 formed shrouded by dust 
$\rm\sim14\,Myr$ ago, at least  $\rm\sim5\%$ of the primordial circumstellar
optically thick discs survives to date.  This time-scale for dust-disc survival seems to 
agree with the disc fractions based on $\rm24\mu\,m$ Spitzer Space Telescope data of Chen et al. 
(\cite{Chen05}) and Low et al. (\cite{Low05}), and the disc-lifetime estimate of Armitage, Clarke 
\& Palla (\cite{Ar03}). As a caveat we note that since discs may dissipate from the inside out, 
the detection rate of dust discs at longer wavelengths is not necessarily the same as that 
implied by shorter wavelength observations, which are sensitive to hotter dust.

\subsection{UCAC2 proper motions}
\label{PM}

Cluster membership of MS, candidate-PMS and RBS stars (Figs.~\ref{fig4} and \ref{fig6})
can be further checked by means of proper motion (PM) data.

Proper motion components in right ascension ($\rm\mu_\alpha\times\cos{(\delta)}$) and declination 
($\rm\mu_\delta$) for stars in the field of NGC\,4755 were obtained in UCAC2\footnote{The Second 
U.S. Naval Observatory CCD Astrograph Catalog (Zacharias et al. \cite{ucac2}), available at 
{\em http://vizier.u-strasbg.fr/viz-bin/VizieR?-source=UCAC2}}. Rather than examining separately 
PM components we use the modulus of the projected PM on the sky 
$\rm\mu(\mas)=\sqrt{\left(\mu_\alpha\times\cos(\delta)\right)^2+\mu_\delta^2}$ (Bica \& Bonatto 
\cite{BiBo2005}). To be consistent with the 2MASS analysis PM data were extracted inside an area 
of 50\arcmin\ in radius centered on the optimized coordinates of NGC\,4755 (Sect.~\ref{n4755}). 
Since UCAC2 also includes 2MASS photometry we verified that the correspondence between both catalogues 
is nearly complete for $7\leq\jj\leq14$. The few bright stars of NGC\,4755 are not included in UCAC2.

\begin{figure}
\resizebox{\hsize}{!}{\includegraphics{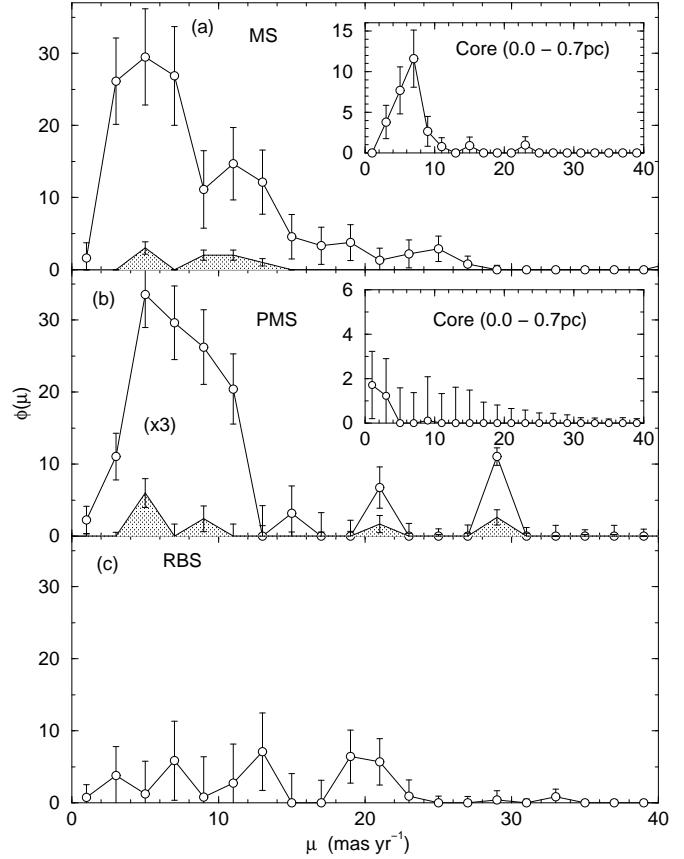}}
\caption[]{Intrinsic proper motion distribution functions of MS (panel a), PMS (panel b) and RBS 
(panel c) stars. Shaded curves: distribution of \ks-excess MS (panel a) and PMS (panel b) stars. 
The spatial region in panels (a) - (c) is $\rm R\leq10\arcmin$  ($\rm R\leq5.3\,pc$). Insets: 
respective PM distributions in the  MS core ($\rm R\leq0.74\,pc$). No \ks-excess star was found 
in the core. The scale in panel (b) was multiplied by 3.}
\label{fig7}
\end{figure}

To isolate the intrinsic PM distribution we take into account the background contamination 
by first applying the MS/evolved, PMS and RBS colour-magnitude filters (Fig.~\ref{fig3}) to the regions 
$\rm r\leq10\arcmin$ (cluster) and $\rm 30\arcmin\leq r\leq50\arcmin$ (offset field). The region 
$\rm r\leq10\arcmin$ or $\approx5.3\,pc$, that matches the limiting radius of NGC\,4755 
(Sect.~\ref{struc}), provides an optimized density contrast between cluster and offset field
(Fig.~\ref{fig5}). Next we build histograms with the number of MS/evolved, PMS and RBS stars in PM 
bins of $2\,\mas$ width, both for cluster and offset regions. Offset field histograms 
are scaled to match the projected cluster area. Finally, the residual field-star 
contamination is statistically eliminated by subtraction of the offset field histogram 
from that of the cluster for MS/evolved, PMS and RBS stars. The intrinsic PM distribution 
functions ($\rm\phi(\mu)=\frac{dN}{d\mu}$) are given in Fig.~\ref{fig7}.

The PM distribution of MS stars (panel a) presents a peak at $\rm\mu\approx5\,\mas$ and 
a secondary one at $\rm\mu\approx11\,\mas$. As discussed in Bica \& Bonatto (\cite{BiBo2005}),
the low-velocity peak is a consequence of the random collective motion of single 
stars superimposed on the cluster's systemic motion. The high-velocity peak is associated with 
unresolved binary (and multiple) systems, where the presence of a secondary changes appreciably 
the velocity of the primary star. This effect was shown for M\,67 in Bica \& Bonatto 
(\cite{BiBo2005}). The binary peak is also present in NGC\,3680 (Bonatto, Bica \& Pavani
\cite{BBP2004}).

Considering error bars, the PM distribution of PMS stars (panel b) shows an excess over the 
field stars for $\rm\mu\leq13\,\mas$ that coincides with the range in PM covered
by the single and binary-star peaks of the MS distribution (panel a). This reinforces cluster 
membership of the candidate-PMS stars.

The PM distribution of RBS stars, on the other hand, represents Poisson fluctuations
of field stars, which agrees with the RDP analysis (Sect.~\ref{struc}).

In Fig.~\ref{fig7} we also analyze PM distributions of \ks-excess MS and PMS stars. Considering
the small number of these objects, we conclude that \ks-excess MS stars share essentially the 
same PM distribution as the combined (with and without \ks-excess) MS single stars. A similar
conclusion applies to the \ks-excess PMS stars. Together with their locus in the field-star
decontaminated CMD (Fig.~\ref{fig4}), PM analysis confirms cluster membership of the \ks-excess 
MS and PMS stars.

The PM distribution of MS stars in the core ($\rm R\leq0.74\,pc$) of NGC\,4755 (inset of panel a) 
presents a low-velocity (single-star) peak. This suggests that the fraction of binaries 
in the core is not significant, as compared with that in the halo.  The inset of panel (b)
shows that the PM distribution of the PMS stars in the core of NGC\,4755 presents evidence of
a low-velocity peak. To avoid inconsistency in spatial regions we only consider the PMS stars 
found in the MS core ($\rm R\leq0.74\,pc$). The large uncertainties preclude further analysis. 
The core contains no \ks-excess MS or PMS stars. 

\section{Luminosity and mass functions}
\label{MF}

To analyze the spatial dependence of LFs and MFs $\left(\phi(m)=\frac{dN}{dm}\right)$ in 
NGC\,4755 we consider the regions: {\em (i)} - core ($\rm 0.0\leq R(pc)\leq0.74$), 
{\em (ii)} halo ($\rm 0.74\leq R(pc)\leq6.94$), and {\em (iii)} overall cluster ($\rm 0.0\leq 
R(pc)\leq6.94$). To maximize the statistical significance of field-star counts we take as the offset 
field the region at $\rm 15\leq R(pc)\leq 25$, that lies $\rm\geq8\,pc$ beyond the limiting 
radius. 

\begin{figure} 
\resizebox{\hsize}{!}{\includegraphics{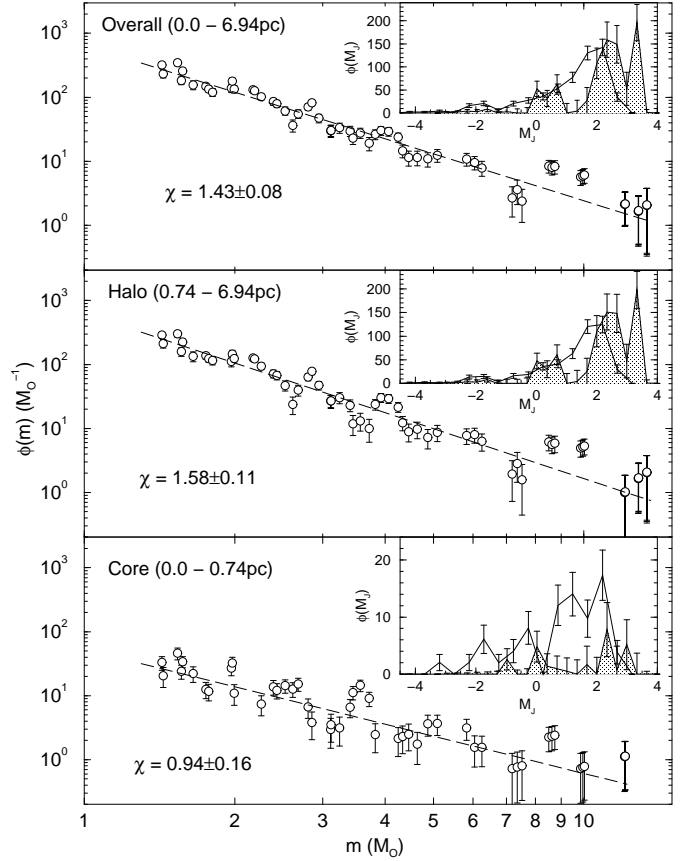}}
\caption[]{Core, halo and overall mass functions of MS stars (empty circles) in NGC\,4755. Each 
panel contains MFs combining \jj, \hh\ and \ks\ 2MASS photometry. MF fits $\left(\phi(m)\propto 
m^{-(1+\chi)}\right)$ are shown as dashed lines. Insets: MS (heavy-solid line) and PMS 
(shaded curve) field-star subtracted LFs. PMS stars are scarce in the core.}
\label{fig8}
\end{figure}

First we isolate MS stars with the colour-magnitude filter (Fig.~\ref{fig3}) 
in the range $7\leq\jj\leq14$. The filtering takes into account most 
of the field stars, leaving a residual contamination. We deal with this statistically by 
building LFs for each cluster region and offset field separately. The faint-magnitude limit of the 
MS is significantly brighter than that of the 99.9\% Completeness Limit 
(Sect.~\ref{2mass}). \jj, H and \ks\ LFs are built by counting stars in magnitude bins from the 
respective faint magnitude limit to the turn-off, for cluster and offset field regions. To avoid 
undersampling near the turn-off and oversampling at the faint limit, magnitude bins are wider in 
the upper MS than in the lower MS. Corrections are made for different solid angles between offset 
field and cluster regions. Intrinsic LFs are obtained by subtracting the respective  
offset-field LFs (Fig.~\ref{fig8}). Finally,
the intrinsic LFs are transformed into MFs using the mass-luminosity relation obtained from the 
14\,Myr Padova isochrone and observed distance modulus $\mMJ=11.5$ (Sect.~\ref{2mass}). These 
procedures are applied independently to the three 2MASS bands. The final MFs are produced by combining 
the \jj, \hh\ and \ks\ MFs. Fig.~\ref{fig8} shows core, halo and overall MFs, covering the mass range 
$\rm1.4\leq m(\ms)\leq13.5$. To these MFs we fit the function 
$\phi(m)\propto m^{-(1+\chi)}$. The fits are shown in Fig.~\ref{fig8}, and MF slopes in 
Table~\ref{tab1}. PMS stars are scarce in the core compared to the halo.

Both the overall and halo MFs are slightly steeper than  Salpeter's (\cite{Salpeter55}) 
IMF ($\chi=1.35$)  with slopes $\chi=1.43\pm0.08$ and $\chi=1.58\pm0.11$. 
However, the core MF is flatter, with $\chi=0.94\pm0.16$. The increase in MF slope from core to 
halo may reflect some amount of mass segregation in the core. We discuss this point further 
in Sect.~\ref{dyna}. 

\subsection{Observed cluster mass}
\label{TM}

Table~\ref{tab1} presents parameters derived from the core, halo and overall MFs. The number 
of evolved stars (col.~2) in each region of NGC\,4755 was obtained by integration of the 
respective field-star subtracted LF ($\rm\jj\leq7$). Multiplying this number by the mass at 
the turn-off ($\rm m\approx13.5\,\ms$) yields an estimate of the mass stored in evolved stars 
(col.~3). PMS stars were isolated by means of the PMS colour-filter (Fig.~\ref{fig3}). 
The number of PMS stars (col.~4) was estimated similarly as for the evolved stars. Based on the 
observed distribution of PMS stars, evolutionary tracks (Fig.~\ref{fig4}) and the fact 
that MFs in general increase in number for the subsolar-mass range, we assume an average 
PMS mass of 1\,\ms. The mass stored in PMS stars is given in col.~5. The observed number of MS stars 
and their corresponding mass (cols.~7 and 8, respectively) were derived by integrating the MF in the mass 
range 1.4--13.5\,\ms. To these we add the corresponding values of number and mass of PMS and evolved 
stars to derive the total number of observed stars (col.~9), observed mass (col.~10), projected mass 
density (col.~11) and mass density (col.~12). The mass locked up in MS/evolved and PMS stars in 
NGC\,4755 is $\rm\sim1150\,\ms$, $\rm\sim16\%$ of this stored in the core. For the PMS stars
we estimate a mass of $\rm\approx285\,\ms$. 
 
\begin{table*}
\caption[]{Parameters derived from MFs and PMS stars of NGC\,4755}
%\tiny
\label{tab1}
\renewcommand{\tabcolsep}{0.37mm}
\renewcommand{\arraystretch}{1.6}
\begin{tabular}{ccccccccccccccccccc}
\hline\hline
&\multicolumn{2}{c}{Evolved}&&\multicolumn{2}{c}{PMS}&&\multicolumn{4}{c}{Observed MS}&
&\multicolumn{4}{c}{$\rm Observed\ MS+PMS+Evolved$}&&$\tau$\\
\cline{2-3}\cline{5-6}\cline{8-11}\cline{13-16}\\
Region&N$^*$&m & &N$^*$&m &&$\chi_{1.4-13}$&&$\rm N^*$&\mobs&&$\rm N^*$&m&$\sigma$&$\rho$\\
 &(stars)&(\ms)&&(stars)&(\ms) &  &&&($10^2$stars)&($10^2\ms$)&& ($10^2$stars)&
  ($10^2\ms$)&($\rm \ms\,pc^{-2}$)&($\rm \ms\,pc^{-3}$)\\
 (1)& (2) & (3) && (4) &(5) &&(6) && (7)& (8) && (9) & (10)&(11)&(12)&&(13)\\
\hline
Core&$1\pm1$ &$14\pm14$ &&$9\pm2$&$9\pm2$&&$0.94\pm0.16$& &$0.4\pm0.1$&$1.2\pm0.4$
&&$0.5\pm0.1$&$1.5\pm0.4$&$85\pm22$&$86\pm23$&&$40\pm9$\\

Halo&$4\pm2$ &$54\pm27$ &&$278\pm83$&$278\pm83$ &&$1.58\pm0.11$& &$2.2\pm0.3$&$6.4\pm1.1$
 &&$5.0\pm0.9$&$9.7\pm1.4$&$6.5\pm0.9$&$0.7\pm0.1$&&---\\
 
Overall& $5\pm2$&$68\pm27$ &&$285\pm85$&$285\pm85$ &&$1.43\pm0.08$& &$2.6\pm0.3$&$8.0\pm1.1$ 
&&$5.5\pm0.9$&$11.5\pm1.4$&$7.6\pm0.9$&$0.8\pm0.1$&&$0.6\pm0.1$\\ 

\hline\hline
\end{tabular}
\begin{list}{Table Notes.}
\item Observed mass range: $\rm1.5-13.4\,\ms$. Col.~6 gives the MF slope derived for MS stars. 
Col.~13: dynamical-evolution parameter $\rm\tau=age/t_{relax}$. 
\end{list}
\end{table*}

The present mass determination for NGC\,4755 is a factor $\rm\approx1.7$ of that of Tadross et al. 
(\cite{Tad2002}) and $\rm\approx0.7$ the mass in NGC\,6611 (Bonatto, Santos Jr. \& Bica \cite{BSJB05}).
The above mass and density estimates (cols.~(10) - (12) in Table~\ref{tab1}) are lower limits, 
since we do not have stars less massive than 1.4\,\ms.

\subsection{Deficiency of PMS stars in the core}
\label{DPMS}

During the early phases of PMS evolution, circumstellar DDEs are subject to 
several dissipation processes (Sect.~\ref{intro}). In principle 
these processes should depend on properties of the central star (mass, radius and luminosity), 
with no strong dependence on position in a star cluster. In this sense, the spatial distributions 
of MS and PMS stars should be similar. However, MS and PMS stars distribute differently through the 
body of NGC\,4755, with the vast majority of PMS stars located in the halo (Figs.~\ref{fig7} and 
\ref{fig8}). The ratio number of PMS to MS stars is $\rm\frac{N_{PMS}}{N_{MS}}\sim0.25$ in 
the core, increasing to $\rm\sim1.2$ in the halo. The apparent deficiency of the core PMS, compared 
to MS stars, may be linked to the denser environment, since the number-density of observed 
$\rm PMS+MS+Evolved$ stars in the core is $\rm\eta_c\sim26\,stars\,pc^{-3}$ while in the halo it drops 
to $\rm\eta_h\sim0.4\,stars\,pc^{-3}$ (Table~\ref{tab1}). Such densities correspond to average distances 
between stars of $\rm\approx0.2\,pc$ in the core and $\rm\approx0.9\,pc$ in the halo. Processes that
could provide additional dissipation mechanisms of PMS DDEs in the core include: 
{\em (i)} stellar density-enhanced radiation pressure, {\em (ii)} tidal disruption 
and {\em (iii)} primordial O stars.

\subsubsection{Radiation pressure}
\label{RP}

Star clusters are not completely virialized systems. However, young
clusters may reach some level of energy equipartition - particularly in the core 
- as suggested by King-like stellar RDPs, mass segregation and advanced dynamical states 
(e.g. Bonatto \& Bica \cite{BB2005}). These conditions apply to NGC\,4755 (Sect.~\ref{dyna}).
Below we assume a cluster virialized state only to derive an estimate of radiation
pressure effects on DDEs.

In thermodynamic equilibrium radiation pressure relates to energy density as $\rm P_r=\frac{1}{3}\ee$. 
Assuming that uniformly distributed stars are the energy source in a spherical region with radius 
in the range ($\rm r_1, r_2$), energy density can be estimated from 
$\rm\ee=\frac{1}{c}\int^{r_2}_{r_1}\eta\mathcal{F}4\pi\,r^2 dr$, where $\eta$ is 
the number-density of stars, $\rm\mathcal{F}=\frac{L}{4\pi\,r^2}$ is the stellar flux at 
a distance $\rm r$, and $\rm c$ is the speed of light. The average radiation pressure in the 
region is $\rm\overline{P_r}\approx\frac{1}{3c}\eta\,\overline{L}\,(r_2-r_1)$, where the average 
luminosity is calculated from the observed MFs (Fig.~\ref{fig8}), 
$\rm\overline{L}=\int L(m)\phi(m)dm/\int\phi(m)dm$, with the mass-luminosity relation $\rm L(m)$ taken 
from the 14\,Myr Padova isochrone. From Table~\ref{tab1} we derive $\rm\overline{L_c}
\approx130\,L_\odot$ and $\rm\overline{L_h}\approx52\,L_\odot$, respectively for core and halo.
We obtain $\rm\overline{P_r}(core)\sim1.2\times10^{-11}\,dyn\,cm^{-2}$, and a core/halo ratio  
$\rm\frac{\overline{P_r}(core)}{\overline{P_r}(halo)}\approx\left(\frac{\eta_c}{\eta_h}\right)
\left(\frac{\overline{L_c}}{\overline{L_h}}\right)\left(\frac{R_c}{R_h}\right)\approx20$, where 
$\rm R_h=\rl-\rc$ is the halo extension.  This suggests that in the core the radiation field may 
be $\sim20\times$ harder to DDEs than the halo because of the uniform force exerted on dust 
particles ($\rm F_r$) by the average radiation pressure. Because the radiation field in a cluster 
is nearly isotropic, the effective force acting on a dust grain is smaller than $\rm F_r$.
The ratio of the (upper limit) radiation force to the gravitational force acting on a spherical 
particle with radius $\rm r_d$ and density $\rm\rho_d$ located at a distance
$\rm r$ from a star of mass $\rm M$ is $\rm F_r/F_G\leq\frac{3P_r\,r^2}{4GM\rho_d\,r_d}$.

 The mean composition of dust grains in young and old discs seems to resemble that of
the solar system comets (as reviewed by Hillenbrand \cite{H05}), $\rm\sim70-80\%$ amorphous
magnesium-rich olivines, $\rm\sim1-10\%$ crystalline forsterite, $\rm\sim10-15\%$ carbons,
$\rm\sim3-5\%$ irons, and trace elements such as silicas. Densities of these elements are
in the range $\rm\rho_d\approx2 - 5\,g\,cm^{-3}$. Water ice with $\rm\rho_d\approx1\,g\,cm^{-3}$
is another common element found in dust grains. In a favourable scenario for DDE disruption, a dust 
particle with $\rm\rho_d=1\,g\,cm^{-3}$ and $\rm r_d=1\,\mu\,m$ located at $\rm r=500\,AU$ (typical 
circumstellar disc radius - Hillenbrand \cite{H05}) from a $\rm M=1\,\ms$\ star will be subject to 
$\rm F_r/F_G\leq0.06$. For smaller disc radii and/or higher grain densities the ratio is even lower. 
Thus, the average radiation pressure in the core of NGC\,4755 has only a marginal effect on circumstellar 
dust particles. 

\subsubsection{Tidal disruption}
\label{BE}

Tidal disruption is effective when the time between binary encounters is significantly shorter than the 
cluster age. For a dust particle located at a distance $\rm r$ from the center of a star, the tidal radius 
due to a second star is $\rm r_t=1.25r$. The effective volume ($\rm\sim\eta_c^{-1}$) occupied by a star in 
the core is $\rm0.04\,pc^3$. Taking $\rm r$ as the typical disc radius, a star with a tidal cross-section 
$\rm\pi\,r_t^2$ traveling at $\rm\approx3\,km\,s^{-1}$ will cover the effective volume in 
$\rm\sim100\,Myr$, about 8 times the cluster age. Thus, binary encounters in the core are not important 
as DDE disruption mechanism. However, a star located at the core radius feels the combined 
tidal pull of all core stars. This increases the tidal radius to $\rm\approx7\,r$ and the time 
between tidal encounters drops to $\rm\sim3\,Myr$. A more realistic estimate must be an intermediate 
value, suggesting that tidal disruption may have played a role in disrupting PMS DDEs
in the core.

\subsubsection{O stars}
\label{Os}

The few SGs of NGC\,4755 (Sect.~\ref{n4755}) evolved from massive stars. We cannot exclude the presence 
of other past massive stars that evolved into compact residuals. An intense radiation field of 
O stars, particularly in the UV, certainly eroded some DDEs in a short period ($\rm\sim4\,Myr$). 

Perhaps the small amount of differential reddening in NGC\,4755 (Sagar \& Cannon \cite{Sagar95}) is a 
relic of dissipated circumstellar envelopes by means of the processes discussed above.

\section{Dynamical state of NGC\,4755}
\label{dyna}

Within uncertainties, the overall MF slope ($\chi=1.43\pm0.08$) in the mass range $\rm1.4\leq 
m(\ms)\leq13.5$ agrees with that of a Salpeter's ($\chi=1.35$) IMF, and MF slopes present 
significant spatial variations, being flatter in the core and rather steep in the halo 
(Table~\ref{tab1} and Fig.~\ref{fig8}). In older clusters this fact reflects dynamical mass 
segregation, in the sense that low-mass stars originally in the core are transferred to the 
cluster's outskirts, while massive stars sink in the core. This process produces a flat core MF 
and a steep one in the halo (e.g. Bonatto \& Bica \cite{BB2005}).

Mass segregation in a star cluster scales with the relaxation time $\rm 
\tr=\frac{N}{8\ln N}\tcr$, where $\rm\tcr=R/\sigma_v$ is the crossing time, N is the (total) number 
of stars and $\rm\sigma_v$\ is the velocity dispersion (Binney \& Tremaine \cite{BinTre1987}). 
Relaxation time is 
the characteristic time-scale for a cluster to reach some level of energy equipartition. For a typical
$\rm\sigma_v\approx3\,\kms$ (Binney \& Merrifield \cite{Binney1998}) we obtain $\rm\tr\approx24\pm5$\,Myr 
for the overall cluster and $\rm\tr\approx0.4\pm0.1$\,Myr for the core. The $\approx14$\,Myr age of NGC\,4755 
(Sect.~\ref{2mass}) corresponds to $\rm\sim40\times\tr(core)$. Thus, some degree of MF slope flattening in 
the core is expected. However, the ratio cluster age 
to \tr\ drops to $\sim0.6$ for the overall cluster, consistent with the Salpeter slope and 
absence of important large-scale mass segregation. 

Bonatto \& Bica (\cite{BB2005}) derived a set of parameters related to the structure and
dynamical evolution of open clusters in different dynamical states. They analysed
nearby open clusters with ages in the range $\rm70-7\,000\,Myr$ and masses in $\rm400-5\,300\,\ms$, 
following the methodology now employed. The evolutionary parameter $\rm\tau=age/\tr$ 
was found to be a good indicator of dynamical state. In particular, significant flattening 
in core and overall MFs due to dynamical effects such as mass segregation is expected to occur 
for $\rm\tau_{core}\geq100$ and $\rm\tau_{overall}\geq7$, respectively.

\begin{figure} 
\resizebox{\hsize}{!}{\includegraphics{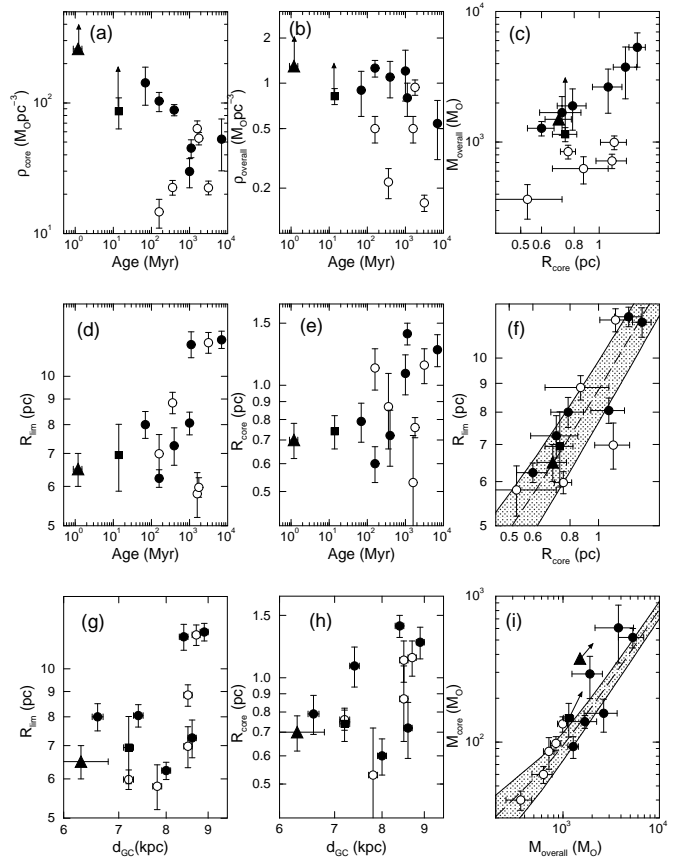}}
\caption[]{Relations involving structural parameters of open clusters. Filled circles:
clusters more massive than 1\,000\,\ms. Open circles, $\rm m<1\,000\,\ms$. Filled triangle:
NGC\,6611. Filled square: NGC\,4755. Dashed lines: least-squares fits to nearby clusters. 
Shaded areas: $\rm1\sigma$ borders of least-squares fits. Arrows indicate 
lower-limit estimates of mass, density and evolutionary parameter for NGC\,4755 and NGC\,6611.}
\label{fig9}
\end{figure}

Fig.~\ref{fig9} focuses on structural parameters, having in mind that mass and density values of 
NGC\,4755 are lower limits (Sect.~\ref{TM}). The limiting radius  is consistent with the 
low-limit correlation of \rl\ with age (panel d). Although with a larger scatter, the same is observed 
in the correlation of \rc\ with age (panel e). The lower-limit of the core density of NGC\,4755 follows 
the trend presented by massive clusters for young ages (panel a). A similar trend is seen for the overall 
density (panel b). In Galactocentric distance the limiting radius of NGC\,4755 helps define a correlation 
(panel g) in the sense that clusters at larger \dgc\ tend to be larger (Lyng\aa\ \cite{Lynga82}; 
Tadross et al. \cite{Tad2002}). The large scatter in panel (h) precludes any 
conclusion with respect to a dependence of core radius with \dgc. NGC\,4755 fits in the tight 
correlations of core and overall mass (panel i), and core and limiting radii (panel f). 
Bonatto \& Bica (\cite{BB2005}) found that massive and less-massive clusters follow different, 
parallel paths in the plane of $\rm core\ radius \times overall\ mass$. NGC\,4755 follows the 
massive cluster relation (panel c). Further details on parameter correlations are given in Bonatto \& Bica 
(\cite{BB2005}).

\begin{figure} 
\resizebox{\hsize}{!}{\includegraphics{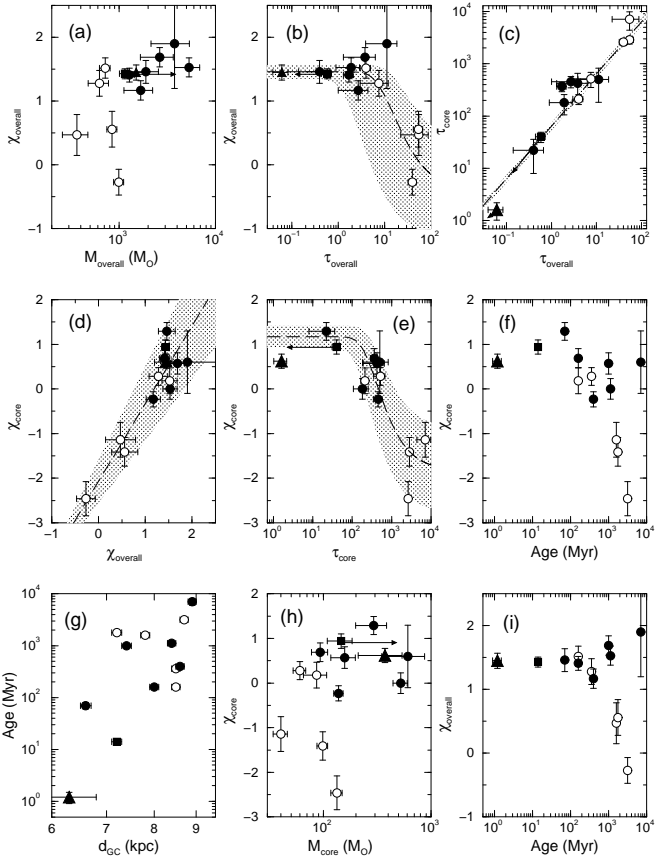}}
\caption[]{Relations involving evolutionary parameters of open clusters. Symbols as in 
Fig.~\ref{fig9}}
\label{fig10}
\end{figure}

Dynamical-evolution parameters of NGC\,4755 are compared with those of other open clusters in 
different dynamical states in Fig.~\ref{fig10}. NGC\,4755 fits in the tight correlations of core 
and overall MF slopes (panel d), and core and overall evolutionary parameters (panel c). The young 
age and rather small Galactocentric distance of NGC\,4755, together with those of NGC\,6611, 
suggest a correlation of \dgc\ with age (panel g) that would agree with Lyng\aa\ (\cite{Lynga82}). 
The Salpeter-like overall MF slope of NGC\,4755 is consistent with the relations for 
$\rm\chi_{overall}\times M_{overall}$ and $\rm\chi_{overall}\times age$ (panels a and i, 
respectively). The scatter in panel (h) does not allow inferences on the relation of 
$\rm\chi_{core}$ with mass. With respect to the evolutionary parameters of NGC\,4755, 
$\rm\chi_{overall}$ follows the relation suggested by massive clusters with small values of 
$\tau$ (panel b). The core MF slope of NGC\,4755 is flatter ($\rm1\sigma$ level) than the expected 
one for its value of $\tau$ (panel e).  Core MF flattening with cluster age is illustrated in
panel (f), where a higher flattening degree in the MFs of old less-massive clusters with respect
to old massive ones occurs that probably reflects a looser stellar distribution. As they age, 
less-massive clusters tend to lose larger fractions of low-mass 
stars through internal processes (e.g. mass segregation and evaporation), and external ones 
(e.g. tidal stripping, disc shocking and encounters with molecular clouds - Bonatto \& 
Bica \cite{BB2005} and references therein). The locus occupied by the core MF of NGC\,4755 
is consistent with the cluster mass and age.

We conclude that both structurally and dynamically, the overall parameters of NGC\,4755 are
consistent with those expected of an open cluster more massive than 1\,000\,\ms\ and 
$\rm\approx14\,Myr$ old. The same applies to core-structural parameters. The rather flat MF in the core, 
on the other hand, cannot result from dynamical mass segregation only, since $\rm\tau_{core}\sim40$ is
rather small. This is similar to what is observed 
in the core of the very young open cluster NGC\,6611 ($\rm\chi_{core}\approx0.6$ and 
$\rm\tau_{core}\approx2$). Primordial conditions associated with fragmentation of the parent molecular 
cloud must have played a role in defining the current core dynamical state. Additional support for this 
scenario is the near absence of binaries in the core compared to the halo of NGC\,4755 (Sect.~\ref{PM}).

\section{Concluding remarks}
\label{Conclu}

In this paper we analyzed the structure, membership and distribution of MS and PMS stars in the 
post-embedded young open cluster NGC\,4755 by means of \jj, \hh\ and \ks\ 2MASS photometry and
UCAC2 proper motions. Field-star decontamination was applied to the photometry to uncover 
the intrinsic CMD morphology. King's model fit to the MS radial density profile produced a core 
radius $\rc=0.74\pm0.08$\,pc with a limiting radius $\rl\approx6.9\pm1.1$\,pc. Similar parameters 
were derived using PMS stars.

Field-star decontaminated CMDs present a well-defined MS with stars more massive than 
$\approx1.4\,\ms$, $\rm\sim285$ candidate-PMS stars, 1 red SG and a few blue SG stars. A 
small fraction of MS and PMS stars present evidence of \ks-excess emission in this 14\,Myr 
cluster. Cluster membership 
of candidate-PMS stars was confirmed by {\it (i)} their locus on the field-star decontaminated 
CMD, {\it (ii)} distribution on the 2-CD, {\it (iii)} cluster-like radial distribution of stars, 
and {\it (iv)} cluster-like proper-motion properties. Membership of \ks-excess MS and PMS stars 
was indicated by arguments {\it (i)} and {\it (iv)}. \ks-excess stars are absent
in the core. From proper motions we found a significant number of binaries in the halo, but
they are scarce in the core.

The halo MF of NGC\,4755 is similar to a Salpeter IMF with a slope $\rm\chi=1.58\pm0.11$,
while in the core it is flatter, $\rm\chi=0.94\pm0.16$. This change in MF slope implies that 
dynamical mass
segregation has affected stellar distribution in the core, since the core-relaxation time 
$\rm\tr(core)\sim0.35$\,Myr corresponds to only $\sim2.5\%$ of the cluster age. For the overall 
cluster $\rm\tr(overall)\sim24\,Myr\ or \sim1.7\times\,$cluster age, which means that mass segregation 
did not have time to redistribute stars on a large scale throughout the halo of such a young cluster.
The MF flattening in the core of NGC\,4755 seems to be related to initial 
conditions, probably associated with the fragmentation of the parent molecular cloud, with more massive 
proto-stars preferentially located in the central parts of the cloud. 

The age of NGC\,4755 was constrained to $\rm14\pm2\,Myr$, consistent with low fractions of 
\ks-excess MS and PMS stars ($\rm f_{K_s}\approx5\%$). The observed mass locked up in MS, PMS 
and evolved stars is $\rm(1.15\pm0.14)\times10^3\,\ms$. Compared to MS stars there is
a deficiency of PMS stars in the core of NGC\,4755. The ratio of PMS to MS stars in the core is 
$\rm\sim1/5$ of that in the halo. Part of this can be accounted for by tidal disruption of 
DDEs in the stellar density-enhanced core, and/or the presence of O stars in
the early phase of the cluster. 

Comparing with theoretical tracks we detected member-PMS stars with ages in the range
$\rm\sim1 - 14\,Myr$. This in turn implies that the star-formation time-scale in NGC\,4755
is at least $\rm10\,Myr$. If the main sources of \ks-excess emission are optically thick 
circumstellar dust discs, detection of \ks-excess in member MS stars indicates that the lifetime 
of some discs may be as large as the age of NGC\,4755. 

\begin{acknowledgements}
 We thank the referee for helpful suggestions.
This publication makes use of data products from the Two Micron All Sky Survey, which 
is a joint project of the University of Massachusetts and the Infrared Processing and 
Analysis Center/California Institute of Technology, funded by the National Aeronautics 
and Space Administration and the National Science Foundation. We also employed the WEBDA 
open cluster database and proper motion data from UCAC2 (The Second U.S. Naval Observatory 
CCD Astrograph Catalog). CB, EB and BB acknowledge support from the Brazilian Institutions 
CNPq and FAPESP. SO acknowledges the Italian Ministero dell'Universit\`a e della Ricerca
Scientifica e Tecnologica (MURST), under the program on 'Fasi Iniziali di Evoluzione dell'Alone 
e del Bulge Galattico' (Italy).
\end{acknowledgements}

%sssssssssssssssssssssssssssss REFERENCESsssssssssssssssssssssssssssssss
%


\begin{thebibliography}{}

\bibitem[1999]{DA99}
   D'Alessio, P., Calvet, N., Hartmann, L., Lizano, S. \& Cant, J. 1999, ApJ, 527, 893
   
\bibitem[2005]{AW05}
   Andrews, S.M. \& Williams, J.P. 2005, ApJ, 631, 1134

\bibitem[2003]{Ar03}
   Armitage, P.J., Clarke, C.J. \& Palla, F. 2003, MNRAS, 342, 1139
   
\bibitem[2005]{BiBo2005}   
    Bica, E. \& Bonatto, C.J. 2005, A\&A, 431, 943
    
\bibitem[2005]{LowC05} 
   Bica, E. \& Bonatto, C.J. 2005, A\&A, 443, 465

\bibitem[1987]{BinTre1987}
   Binney, J., \& Tremaine, S. 1987, in {\it Galactic Dynamics}, Princeton,
   NJ: Princeton University Press. (Princeton series in astrophysics)

\bibitem[1998]{Binney1998}
   Binney, J., \& Merrifield, M. 1998, in {\it Galactic Astronomy}, Princeton,
   NJ: Princeton University Press. (Princeton series in astrophysics) 
   
\bibitem[2006]{BSJB05} 
   Bonatto, C.J., Santos Jr., J.F.C. \& Bica, E. 2006, A\&A, 445, 567

\bibitem[2005]{BB2005} 
   Bonatto, C.J., \& Bica, E. 2005, A\&A, 437, 483.
   
\bibitem[2005]{BBS2005} 
   Bonatto, C.J., Bica, E., \&  Santos Jr., J.F.C. 2005, A\&A, 433, 917.
   
\bibitem[2004]{BBP2004} 
   Bonatto, C.J., Bica, E., \& Pavani, D.B. 2004, A\&A, 427, 485
   
\bibitem[2004]{BBG2004} 
   Bonatto, C., Bica, E., \& Girardi, L. 2004, A\&A, 415, 571

\bibitem[2003]{BB2003} 
   Bonatto, C., \& Bica, E. 2003, A\&A, 405, 525     

\bibitem[2000]{Brand00} 
   Brandner, W., Zinnecker, H., Alcalá, J.M. et al. 2000, AJ, 120, 950
   
\bibitem[2005]{Chen05}
   Chen, C.H., Jura, M., Gordon, K.D. \& Blaylock, M. 2005, ApJ, 623, 493

\bibitem[1984]{DK84}
   Dachs, J. \& Kaiser, D., 1984, A\&AS, 58, 411   
  
\bibitem[2002]{DSB2002} 
   Dutra, C.M., Santiago, B.X., \& Bica, E. 2002, A\&A, 381, 219  
     
\bibitem[2002]{Girardi2002} 
   Girardi, L., Bertelli, G., Bressan, A., et al. 2002, A\&A, 391, 195  
 
\bibitem[2005]{H05}
   Hillenbrand, L.A. 2005, in {\em A Decade of Discovery: Planets Around Other Stars},
   STScI Symposium Series 19, ed. M. Livio, in press (astro-ph/0511083)

\bibitem[1995]{KH95}
   Kenyon, S.J. \& Hartmann, L. 1995, ApJS, 101, 117
  
\bibitem[1966a]{King1966a}
   King, I. 1966a, AJ, 71, 64
   
\bibitem[1966b]{King1966b}
   King, I. 1966b, AJ, 71, 276  

\bibitem[2002]{Kroupa2002}
   Kroupa, P. 2002, Science, 295, 82
   
\bibitem[2004]{Kroupa2004}
   Kroupa, P. 2004, New Astron. Rev., 48, 47
   
\bibitem[1992]{LaAd92} 
   Lada, C.J. \& Adams, F.C. 1992, ApJ, 393, 278

\bibitem[2005]{Low05}
   Low, F.J., Smith, P.S., Werner, M., Chen, C.H., Krause, V., Jura, M. \& Hines, D.C. 
   2005, ApJ, 631, 1170
   
\bibitem[1982]{Lynga82}
   Lyng\aa, G. 1982, A\&A, 109, 213

\bibitem[1987]{Lynga87}
   Lyng\aa, G. 1987, in {\em Catalog of Open Cluster Data}, Computer Based
   Catalogue available through the CDS, Strasbourg, France and NASA DATA
   Center, Greenbelt, Maryland, USA, 5th edition, 4, 121
     
\bibitem[1996]{Merm1996}   
   Mermilliod, J.C. 1996, in {\it The Origins, Evolution, and Destinies of 
   Binary Stars in Clusters}, ASP Conference Series, eds. E.F. Milone \& J.-C. 
   Mermilliod, 90, 475

\bibitem[2002]{Nilakshi2002}
   Nilakshi, S.R., Pandey, A.K. \& Mohan, V. 2002, A\&A, 383, 153

\bibitem[2005]{OBBM05}
   Ortolani, S., Bica, E., Barbuy, B. \& Momany, Y. 2005, in {\em Protostars and Planets V},
   Ed. B. Reipurth, Kona, Hawaii

\bibitem[2004]{Piskunov04}
   Piskunov, A.E., Belikov, A.N., Kharchenko, N.V., Sagar, R. \& Subramaniam, A. MNRAS, 349, 1449
   
\bibitem[1993]{Reid93}
   Reid, M.J. 1993, ARA\&A, 31, 345 

\bibitem[1995]{Sagar95}
   Sagar, R. \& Cannon, R.D. 1995, A\&AS, 111, 75

\bibitem[2001]{SBWG01}
   Sanner, J., Bruzendorf, J., Will, J.-M. \& Geffert, M. 2001, A\&A, 369, 511
   
\bibitem[1982]{SK82}
   Schmidt-Kaler, T. 1982, in {\it Landolt-B\"ornstein, New Ser., Group VI}, vol. 
   2b (Springer-Verlag, Berlin), 1
   
\bibitem[1955]{Salpeter55}
   Salpeter, E. 1955, ApJ, 121, 161
   
\bibitem[2000]{Siess2000}
   Siess, L., Dufour, E., \& Forestini, M. 2000, A\&A, 358, 593

\bibitem[1997]{2mass1997} 
   Skrutskie, M., Schneider, S.E., Stiening, R., et al. 1997, in {\it The Impact 
   of Large Scale Near-IR Sky Surveys}, ed. Garzon et al., Kluwer (Netherlands), 210, 187
   
\bibitem[2002]{Tad2002}
   Tadross, A.L., Werner, P., Osman, A. \& Marie, M. 2002, NewAst, 7, 553

\bibitem[1995]{TKD95}
   Trager, S.C., King, I.R., \& Djorgovski, S. 1995, AJ, 109, 218

\bibitem[2005]{WH05}
   White, R.J. \& Hillenbrand, L.A. 2005, ApJL, 621, 65

\bibitem[2003]{ucac2}
   Zacharias, N., Urban, S.E., Zacharias, M.I., Wycoff, G.L., Hall, D.M.,
   Monet, D.G., \& Rafferty, T.J.  2004, AJ, 127, 3043   
   
\end{thebibliography}
\end{document}